\documentclass[12pt]{article}

\usepackage{bbm}
\usepackage{epsfig}
\usepackage{array}
\usepackage{float}

\usepackage{a4}
\usepackage{a4wide}
\def\gs{\mathrel{
   \rlap{\raise 0.511ex \hbox{$>$}}{\lower 0.511ex \hbox{$\sim$}}}}
\def\ls{\mathrel{
   \rlap{\raise 0.511ex \hbox{$<$}}{\lower 0.511ex \hbox{$\sim$}}}}

\newcommand{\obb}{0\mbox{$\nu\beta\beta$}}
\newcommand{\onbb}{neutrino-less double beta decay}
\newcommand{\ba}{\begin{array}{c}}
\newcommand{\baz}{\begin{array}{cc}}
\newcommand{\bad}{\begin{array}{ccc}}
\newcommand{\bav}{\begin{array}{cccc}}
\newcommand{\bea}{\begin{equation} \begin{array}{c}}
\newcommand{\eea}{ \end{array} \end{equation}}
\newcommand{\ea}{\end{array}}

\newcommand{\dms}{\mbox{$\Delta m^2_{\odot}$}}
\newcommand{\dma}{\mbox{$\Delta m^2_{\rm A}$}}
\newcommand{\meff}{\mbox{$\left| m_{ee} \right|$}}

\newcommand{\be}{\begin{eqnarray}}
\newcommand{\ee}{\end{eqnarray}}

\begin{document}

\begin{titlepage}
\title{\vspace*{-2.0cm}
%\hfill {\small hep--ph/xxxxxx}\\[20mm]
\bf\Large
Statistical Analysis of future Neutrino Mass Experiments 
including Neutrino-less Double Beta Decay
\\[5mm]\ }

\author{
Werner Maneschg\thanks{email: \tt werner$.$maneschg@mpi-hd.mpg.de}~~,~~
Alexander Merle\thanks{email: \tt alexander$.$merle@mpi-hd.mpg.de}~~,~~
Werner Rodejohann\thanks{email: \tt werner$.$rodejohann@mpi-hd.mpg.de}
\\ \\
{\normalsize \it Max-Planck-Institut f\"ur Kernphysik,}\\
{\normalsize \it Postfach 10 39 80, 69029 Heidelberg, Germany}
}
\date{\today}

\maketitle
\thispagestyle{empty}

%PACS: 14.60.Pq (Neutrino mass and mixing), 23.40.-s (Double $\beta$ decay)

\begin{abstract}
\noindent
We perform a statistical analysis with the prospective results 
of future experiments on neutrino-less double beta decay, 
direct searches for neutrino mass (KATRIN) and cosmological
observations. Realistic errors are used and the nuclear 
matrix element uncertainty for neutrino-less double beta decay 
is also taken into account. 
Three benchmark scenarios are introduced, corresponding to 
quasi-degenerate, inverse hierarchical neutrinos, and  
an intermediate case. We investigate to what extend these scenarios can be
reconstructed. Furthermore, we check the compatibility of 
the scenarios with the claimed evidence of 
neutrino-less double beta decay.
\end{abstract}

\end{titlepage}

%%%%%%%%%%%%%%%%%%%%%%%%%%%%%%%%%%%%%%%%%%%%%%%%%%%%%%%%%%%%%%%%%%%%%%
\section{\label{sec:intro} Introduction}
%%%%%%%%%%%%%%%%%%%%%%%%%%%%%%%%%%%%%%%%%%%%%%%%%%%%%%%%%%%%%%%%%%%%%%
Neutrino mass and lepton mixing represent an unambiguous proof 
that the Standard Model (SM) of elementary particles is incomplete. 
Various experiments with solar \cite{sol}, atmospheric \cite{atm} 
and man-made \cite{kl,minos} neutrino sources imply 
non-trivial lepton mixing angles, as well as non-zero and 
non-degenerate neutrino masses. 
Their values are extremely suppressed with respect to the 
masses of the other (electrically charged) fermions of 
the SM. 
%{\bf 
The most prominent and often studied 
mechanism to explain the smallness of 
neutrino masses is the see-saw mechanism \cite{seesaw}. 
The neutrino mass scale is here  
inversely proportional to the scale of its origin. 
In addition, lepton number violation is predicted: 
neutrinos are Majorana particles. Searching for this property  
will be a crucial test of the see-saw mechanism, but also of other 
mechanisms leading to small Majorana neutrino masses. 
%} 
Possible phenomenological consequences of lepton number 
violation are the generation of the baryon asymmetry of the Universe 
\cite{lepto} or, at low energies, \onbb~(\obb) \cite{APS}. This 
decay of certain nuclei, $(A,Z) \rightarrow (A,Z+2) + 2 \, e^-$,
which has not yet been observed, clearly violates lepton 
number by two units, and is intensively 
searched for \cite{APS}. We will assume here that 
light Majorana neutrinos are exchanged in the diagram responsible 
for \obb. In this case, the amplitude for this process is 
proportional to the coherent sum  
\be \label{eq:meff}
m_{ee}  \equiv 
\sum\limits_{i=1}^3 U_{ei}^2 \, m_i \,,
\ee
where $m_i$ are the individual neutrino masses and $U$ is the 
leptonic mixing, or Pontecorvo-Maki-Nakagawa-Sakata (PMNS), matrix. 
The absolute value of $m_{ee}$ is called the effective mass. 
The entries $U_{ei}$ can be written as 
%\bea \label{eq:PMNS}
$U_{e1} = \cos \theta_{12} \, \cos \theta_{13}$, 
$U_{e2} = \sin \theta_{12} \, \cos \theta_{13} \, e^{i \alpha}$ 
and $U_{e3} = \sin \theta_{13} \, e^{i \beta}$, 
%\eea
where $\alpha$ and $\beta$ are two currently 
unknown ``Majorana phases'' and 
$\theta_{12, 13}$ are mixing angles. While $\theta_{13}$ 
is constrained mainly by short-baseline reactor experiments, 
$\theta_{12}$ is probed by solar and long-baseline reactor 
neutrino experiments. 
Their current best-fit values as well as 
$1\sigma$ and $3\sigma$ ranges can be 
obtained from three-flavor fits, 
the result being \cite{data} 
\be \label{eq:data}
\sin^2 \theta_{12}
 = 0.32\,(\pm 0.02)^{+0.08}_{-0.06} ~,~
\sin^2 \theta_{13} = 0^{+0.019, \, 0.050}\,.
\ee 
In what regards the neutrino masses, 
for a normal ordering one has 
$m_3 > m_2 > m_1$ with $m_2^2 = m_1^{2} +\dms$ and 
$m_3^2 = m_1^{2}+\dma$. In case of an inverted ordering 
we have $m_2 > m_1 > m_3$ with $m_2^2 = m_3^{2}+\dms+\dma$ and 
$m_1^2 = m_3^{2} + \dma$. Here $\dms$ and $\dma$ are  
mass-squared differences with best-fit values and 
$3\sigma$ ranges $(7.9^{+1.1}_{-0.9}) \cdot 10^{-5}$ eV$^2$ and 
$(2.6^{+0.6}_{-0.6}) \cdot 10^{-3}$ eV$^2$, 
respectively \cite{data}.  
Quasi-degenerate neutrino masses occur when 
$m_{1,2,3}^2 \gg \dma, \dms$. 
If neutrinos are Majorana particles, 
all low energy neutrino phenomenology can be described by  
the neutrino mass matrix 
$m_\nu = U^\ast \, m_\nu^{\rm diag} \, U^\dagger$. 
It contains nine physical parameters. 
%Comparing this matrix with the effective mass in eq.~(\ref{eq:meff}) 
%reveals that $m_{ee}$ is the 11-entry of $m_\nu$. Therefore, 
%\onbb~provides the unique opportunity to directly measure an element 
%of the mass matrix. Furthermore, 
Seven out of the nine parameters 
of the neutrino mass matrix appear in \meff. Therefore, it contains 
a large amount of information, in particular if complementary 
measurements of some of the other parameters exist. We also note that 
all parameters of $m_\nu$ which do {\it not} influence 
neutrino oscillations show up in the 
effective mass. Those are the the Majorana phases 
and, in particular, the individual neutrino masses   
(neutrino oscillations are only sensitive to 
mass-squared differences). For a review on 
the dependence of \meff~on the 
various neutrino parameters see 
refs.~\cite{APS,Lindner:2005kr,0vbb_rev} and 
references therein. 
In the present paper, 
in contrast to other works statistically analyzing future neutrino 
mass measurements including 
\obb~\cite{PPS,WR_old,HP,AdG,steen2,Fogli:2006yq,nme1},
we focus on the neutrino mass scale, 
i.e.\ the value of the smallest neutrino mass. 
To this end we define three 
natural benchmark scenarios and investigate how 
future experiments may be able to constrain 
them. Our goal here is to combine as much mass-related 
information as possible.

\section{Observables related to neutrino mass}
\begin{figure}
\epsfig{file=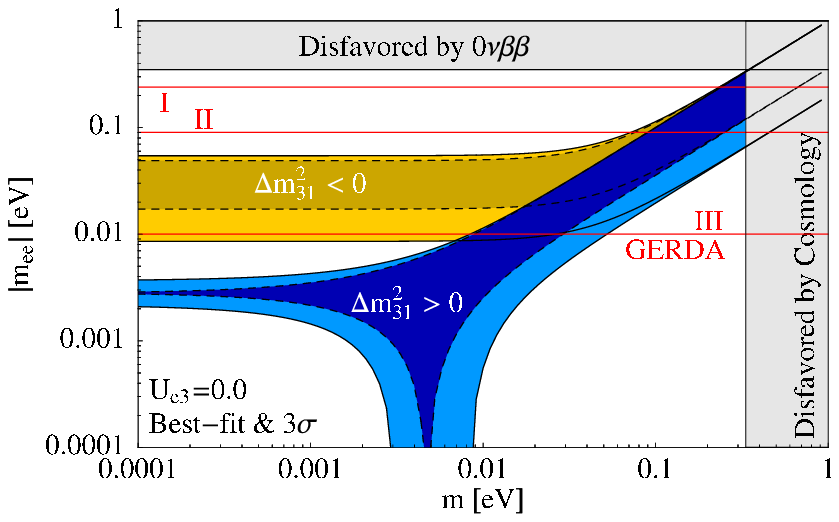,width=7.48cm,height=4.25cm}
\epsfig{file=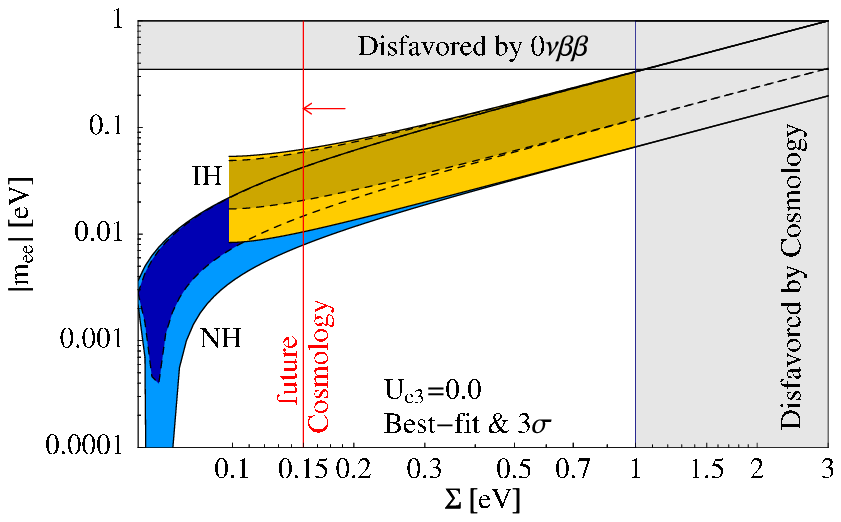,width=7.48cm,height=4.25cm}
\epsfig{file=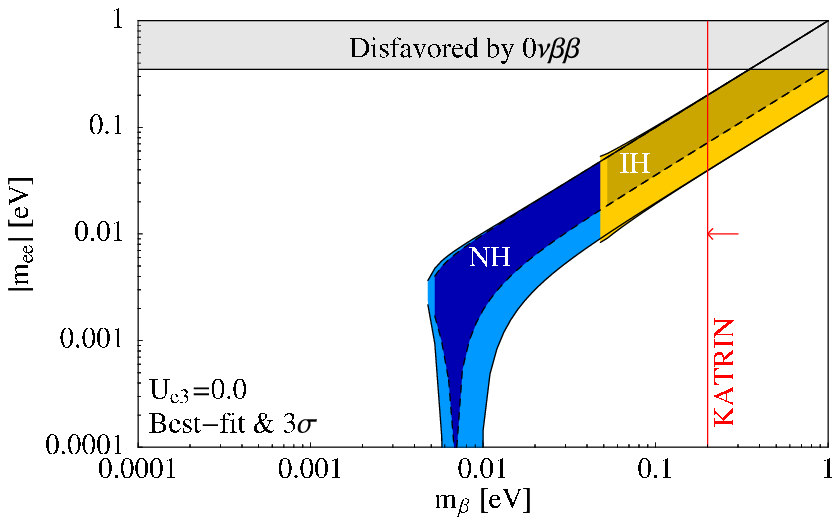,width=7.48cm,height=4.25cm}
\caption{\label{fig:1}The effective mass as a function of 
the smallest neutrino mass, the sum of neutrino masses $\Sigma$ 
and the kinematic neutrino mass $m_\beta$. The value 
$U_{e3} = 0$ and the current 3$\sigma$ 
ranges of the other oscillation parameters have been used.}
\end{figure}

Currently the strongest experimental limits\footnote{We note 
that there is a claimed positive signal for \obb~from 
ref.~\cite{ev}. We will turn to this issue later on.} on the 
half-life of \onbb~are (all at 90 $\%$ C.L.) 
$1.9 \cdot 10^{25}$~y for 
$^{76}$Ge \cite{KK} (see also \cite{igex}), 
T$_{1/2} \ge 3.0 \cdot 10^{24}$~y for $^{130}$Te \cite{cuore}, 
T$_{1/2} \ge 5.8 \cdot 10^{23}$~y for $^{100}$Mo
and T$_{1/2} \ge 2.1 \cdot 10^{23}$~y for $^{82}$Se \cite{nemo}. 
The existing limits on T$_{1/2}$ will be 
improved considerably (by two orders of magnitude or more) 
in the near future by various experiments \cite{APS}.
The uncertainty in nuclear matrix element (NME) 
calculations is a serious problem to translate these bounds into upper limits on the effective mass \cite{nme,nme1}. We will take into account in particular this uncertainty in our analysis. Depending on the nuclei and NME, the current limit on the effective mass as extracted from the 
half-lifes given above lies between several tenths of and a few eV. This has to be compared with the predictions which can be made for the effective mass. Inserting the known ranges of the oscillation 
parameters, and varying the unknown parameters within 
their allowed ranges, one can generate plots as the ones in fig.~\ref{fig:1}. 

They display (for $U_{e3}=0$) the effective mass as a 
function of the smallest neutrino 
mass, the sum of neutrino masses 
\be \label{eq:sigma}
\Sigma \equiv \sum\limits_{i=1}^3 m_i 
\ee
and the kinematic neutrino mass 
\be\label{eq:mbeta}
m_\beta \equiv \sqrt{\sum\limits_{i=1}^3 |U_{ei}|^2 \, m_i^2}\,.
\ee
The latter two quantities can 
be measured through cosmological observations \cite{cosmo} and 
experiments like KATRIN \cite{katrin}, respectively. 
The latter experiment has a $5\sigma$ discovery potential of 0.35 eV 
for $m_\beta$, and a null result will lead to a 90 $\%$ C.L.~limit 
of 0.2 or 0.17 eV \cite{fre_bas}. In the sensitivity range of 
KATRIN, the relation $3\,m_\beta = \Sigma$ holds to a very 
good precision. Cosmology is expected to probe values 
of $\Sigma$ down to the 0.1 eV range 
\cite{cosmo} (to be specific, we take a value of 
0.15 eV in fig.~\ref{fig:1}). To achieve such impressive results, one
takes advantage of 
future observations of weak gravitational lensing of galaxies, and 
the cosmic microwave background or detailed analyses of 
the 21 cm hydrogen emission lines at high redshift. 
It is fair to say that a conservative limit on $\Sigma$ is 1 eV. 
This value corresponds roughly to the bound obtained from WMAP 
5-year data alone \cite{wmap5}.  Recall that neutrino mass 
bounds from cosmology depend strongly on the data sets, the 
priors and the model, i.e., adding parameters which are degenerate
with neutrino masses will relax the bounds, see, e.g., \cite{steen}. 
Finally, current limits for $m_\beta$ are 2.3 eV \cite{mainz}.

The blue and yellow bands in fig.~\ref{fig:1} 
correspond to the normal and inverted mass ordering of 
the neutrinos, respectively. 
The darker areas in the blue and yellow 
bands are obtained when the oscillation parameters are fixed to 
their best-fit values and only the Majorana phases are varied. 
The lighter areas 
correspond to the $3\sigma$ ranges of the oscillation parameters.
Note that this broadening is very weak for the maximum value of 
\meff~in the case of inverted mass ordering and for quasi-degenerate 
neutrinos. This is because the upper limits on \meff~are roughly 
$\sqrt{\dma}$ and $m_3$, respectively, and varying the oscillation 
parameters has very little impact. In the first plot of
fig.~\ref{fig:1}, we have indicated three special values of 
\meff~which correspond to the goals of the three phases of the GERDA experiment (where a certain NME has been assumed, see 
\cite{GERDAproposal} for details).

\section{\label{sec:2}Statistical analysis}
Now we will perform a statistical analysis to investigate how 
well it will be possible to reconstruct different realistic 
physical scenarios with upcoming neutrino mass experiments. 
Note that, since we want to investigate realistic situations, 
we concentrate only on cases that can be probed 
in the near future. For 
definiteness, we consider the inverted mass ordering 
and three different 
scenarios called ${\cal QD}$ (quasi-degenerate), 
${\cal INT}$ (intermediate) and ${\cal IH}$ (inverted hierarchy) 
that are defined by different values of the smallest 
neutrino mass $m_3$. Note that the ${\cal QD}$ scenario would, 
to very large extent, also apply to a normal mass ordering. 
The hypothetical ``true values'' 
for the different observables in these scenarios are:

\begin{table}[h]
\begin{center}
\begin{tabular}{c|cccc}
 Scenario &  $m_3$ [eV] & $\meff$ [eV] & $m_\beta$ [eV] & $\Sigma$
[eV] \\ \hline
 ${\cal QD}$  &  0.3    & $0.11-0.30$    &  0.30   &  0.91\\
 ${\cal INT}$ &  0.1    & $0.04-0.11$    & (0.11)  &  0.32\\
 ${\cal IH}$  &  0.003  & $0.02-0.05$    & (0.05)  & (0.10)
\end{tabular}
\end{center}
\end{table}%\vspace{-.42cm}

%{\bf 
We have used here the best-fit values for the oscillation parameters. 
The range for \meff\ originates from the variation of the 
Majorana phases $\alpha$ and $\beta$.
%} 
Note that the KATRIN experiment will only 
be able to measure $m_\beta$ in the case of the 
${\cal QD}$ scenario, while for ${\cal INT}$ and 
${\cal IH}$ it will only provide an upper limit. 
The same is true for the measurement of $\Sigma$ in the 
${\cal IH}$ scenario. These cases are indicated in the table 
by writing the respective values in brackets. 

Let us now give a summary of the different experimental errors and 
theoretical uncertainties. Regarding the error on the 
effective mass in $0\nu\beta\beta$, we have to distinguish 
between experimental and ``theoretical uncertainties'', 
where the latter result from the NME uncertainty. The experimental 
error can be included by noting that the decay width depends
quadratically on the effective mass. Thus, 
\be
\sigma(\meff_{\rm exp}) = 
\frac{|m_{ee}|_{\rm exp}}{2} \, 
\frac{\sigma(\Gamma_{\rm obs})}{\Gamma_{\rm obs}} \, ,
\label{eq:mee_obs-error}
\ee
where $|m_{ee}|_{\rm exp}$ is the measured value of the 
effective neutrino mass and 
$\sigma(\Gamma_{\rm obs})$ is the experimental error on 
the measured decay width 
$\Gamma_{\rm obs}$ for \onbb. For definiteness, we 
choose the ratio of the latter two as 
\be
\frac{\sigma(\Gamma_{\rm obs})}{\Gamma_{\rm obs}} \simeq  23.3\%\,,
\label{eq:sG_exp}
\ee
which is the value obtainable in the GERDA experiment 
\cite{GERDAproposal}. We combine, similarly to the procedure 
developed in ref.~\cite{WR_old}, the experimental error  
with the theoretical NME error via 
\be
\sigma(\meff) = (1 + \zeta) \, \left(\meff + \sigma(\meff_{\rm exp}) 
\right) - \meff \, ,
\label{eq:mee_err}
\ee
where $\zeta \ge 0$ parameterizes the NME uncertainty and 
$\sigma(\meff_{\rm exp})$ is given in eq.~(\ref{eq:mee_obs-error}). 
Following ref.~\cite{PPS}, we define a covariance 
matrix 
\begin{equation}
 S_{ab}\equiv \delta_{ab} \, \sigma^2 (a) + \sum_i 
\frac{\partial T_a}{\partial x_i} \, 
\frac{\partial T_b}{\partial x_i} \, \sigma^2_i \,,
 \label{eq:cov-matrix}
\end{equation}
where $T_1 = \meff$, $T_2 = \Sigma$ and $T_3 = m_\beta^2$. 
%{\bf 
Furthermore, $\sigma^2 (a)$ is the error on $T_a$, and 
$a$, $b$ label the entries in the covariance matrix. 
The $x_i$ are the oscillation parameters that enter \meff~(and 
$m_\beta$, though in the observable range of $m_\beta$ they have
basically no influence).%} 
The errors on the $T_a$ are given by eq.~(\ref{eq:mee_err}) 
as well as by $\sigma(m_\beta^2) = 0.025$~ eV$^2$~\cite{katrin,fre_bas} 
and $\sigma(\Sigma) = 0.05$~eV~\cite{cosmo}.

Defining $v_a = T_a - (T_a)_{\rm exp}$, where $(T_a)_{\rm exp}$ 
denotes the experimental value of $T_a$, our $\chi^2$-function 
to be minimized is
\begin{equation} \nonumber 
 \chi^2 = v^T  \, S^{-1} \, v\, .
\label{eq:chi2-function}
\end{equation}
%{\bf 
All oscillation parameters are set to their current 
best-fit values and their (symmetrized) standard deviations are 
determined from their 1$\sigma$-ranges, 
which is a good approximation for future 3$\sigma$-ranges. 
Anyway, the impact of different numerical values here would 
not lead to qualitatively different results.%}
We first minimize the $\chi^2$ from eq.~(\ref{eq:chi2-function}) 
with respect to the Majorana phases $\alpha$ and $\beta$. 
The resulting function is 
$\chi^2_{\rm res}=\min_{\alpha,\beta} \chi^2$. 
We then continue by plotting the resulting 
$1\sigma$, $2\sigma$ and $3\sigma$ ranges for the 
smallest neutrino mass $m_3$ determined by setting 
$\Delta \chi^2 = \chi^2_{\rm res}-\chi^2_{\rm res,min} $ 
equal to 1, 4 and 9. This corresponds to a $\chi^2$-function 
with one free parameter (namely $m_3$). $|m_{ee}|_{\rm exp}$ is 
the assumed measured value of \meff, on which the reconstructed 
range of $m_3$ depends. The minimum in the $|m_{ee}|_{\rm exp}$-$m_3$
plane is determined such that $\Delta \chi^2$ is zero in the true
region of the corresponding scenario (e.g., $\cal QD$).
\begin{figure}[t]
\begin{center}
\epsfig{file=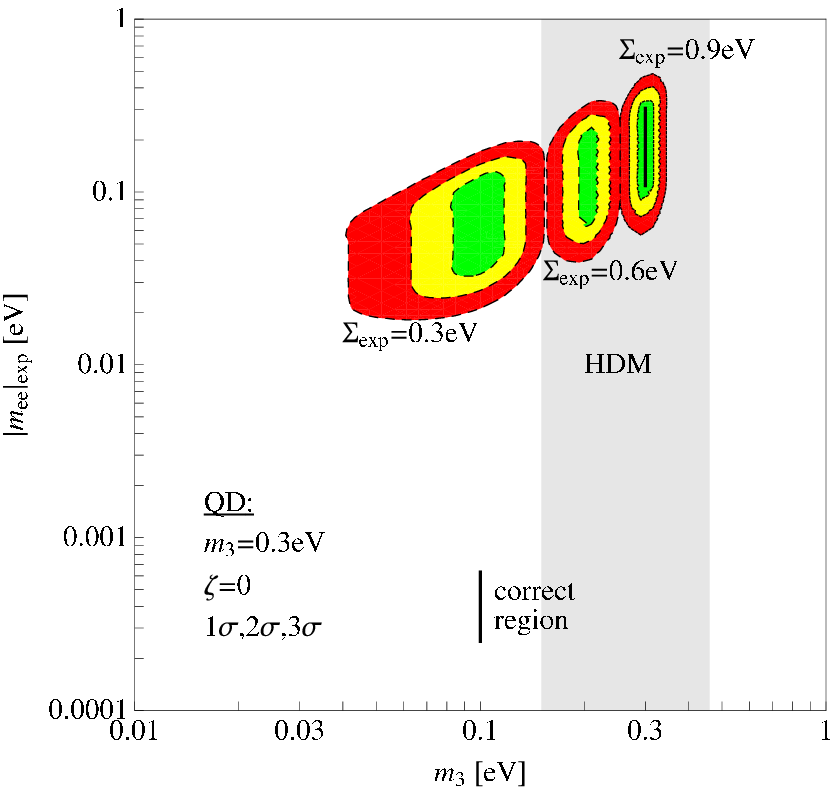,height=5.35cm,width=5.2cm}
\epsfig{file=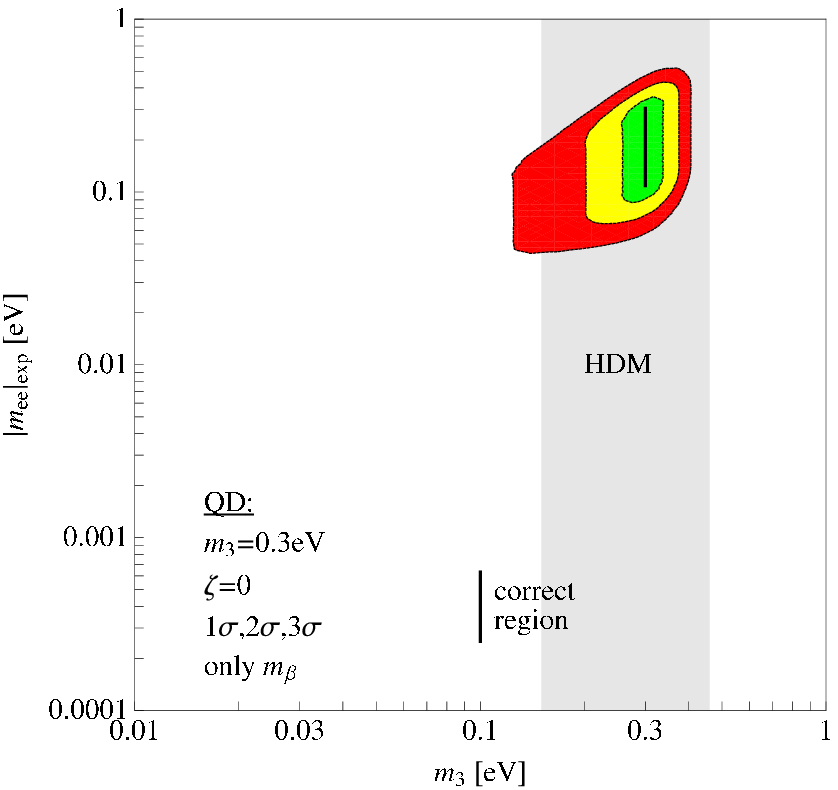,height=5.35cm,width=5.2cm}
\epsfig{file=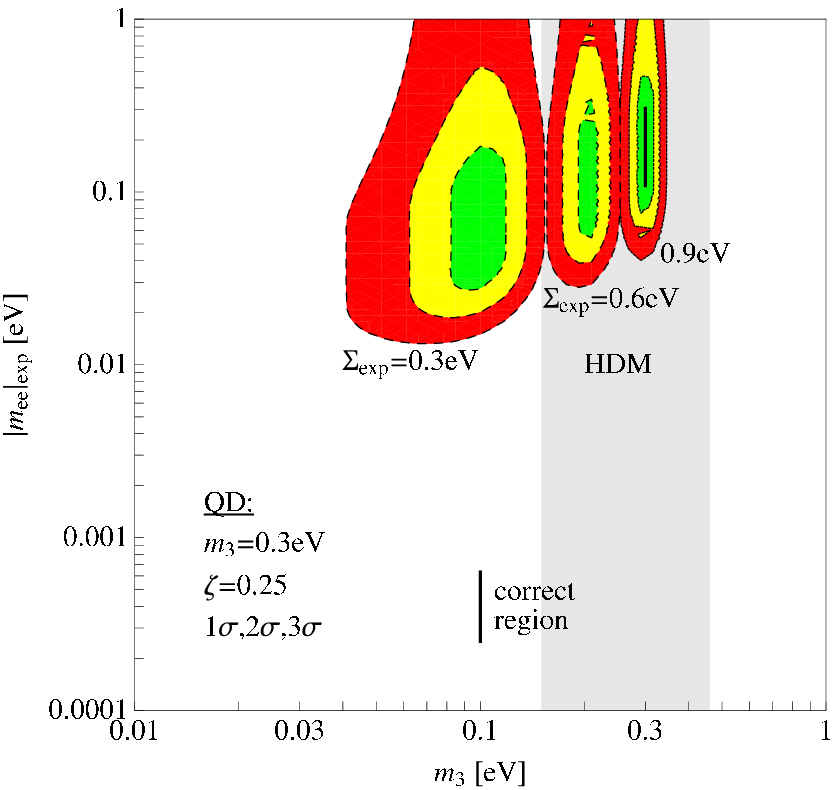,height=5.35cm,width=5.2cm}
\epsfig{file=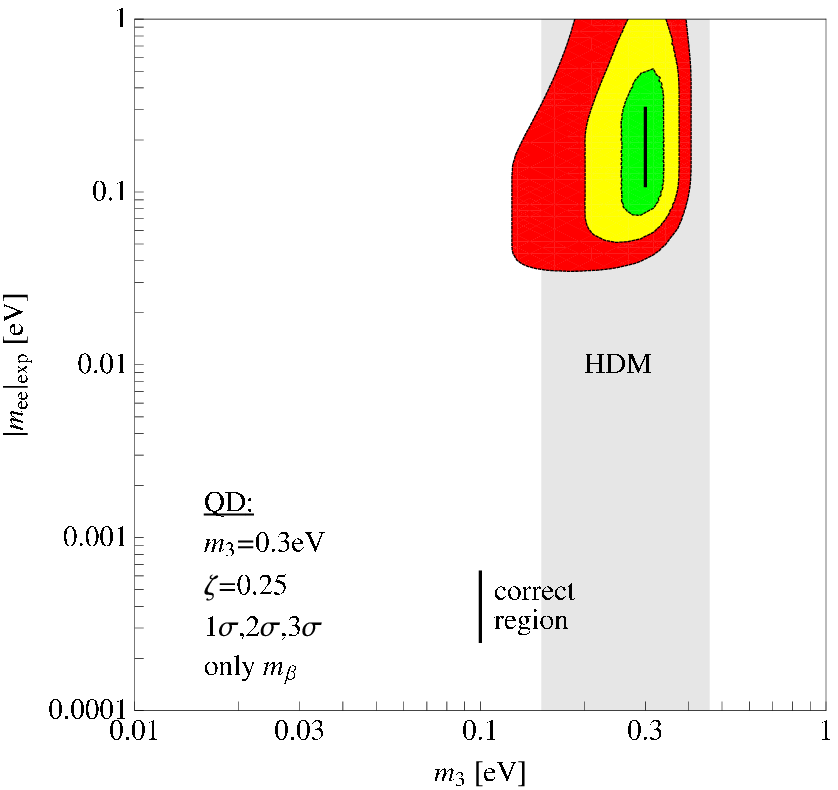,height=5.35cm,width=5.2cm}
\epsfig{file=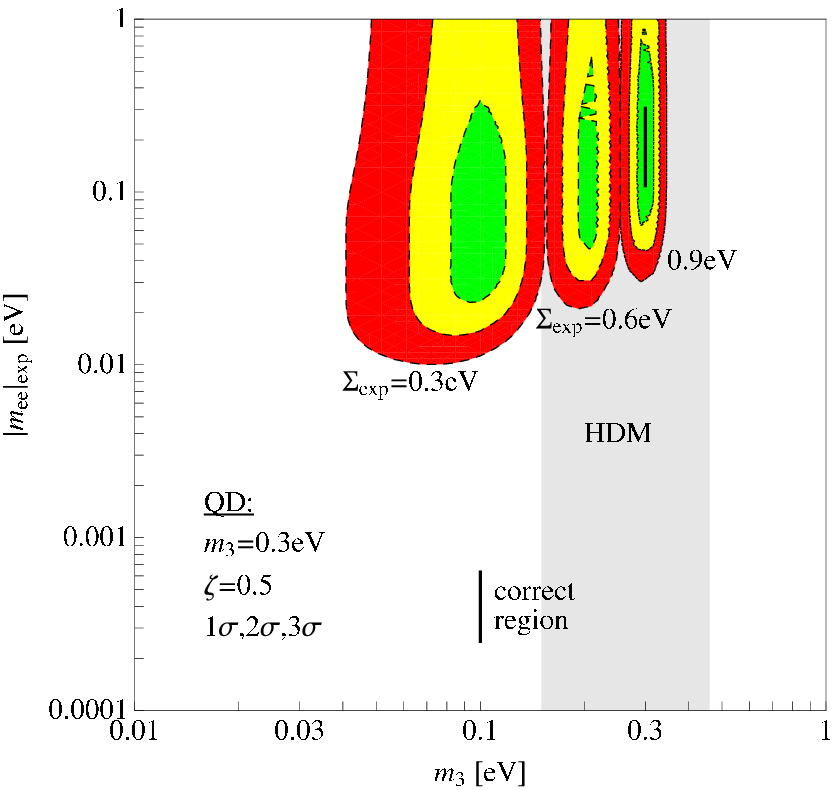,height=5.35cm,width=5.2cm}
\epsfig{file=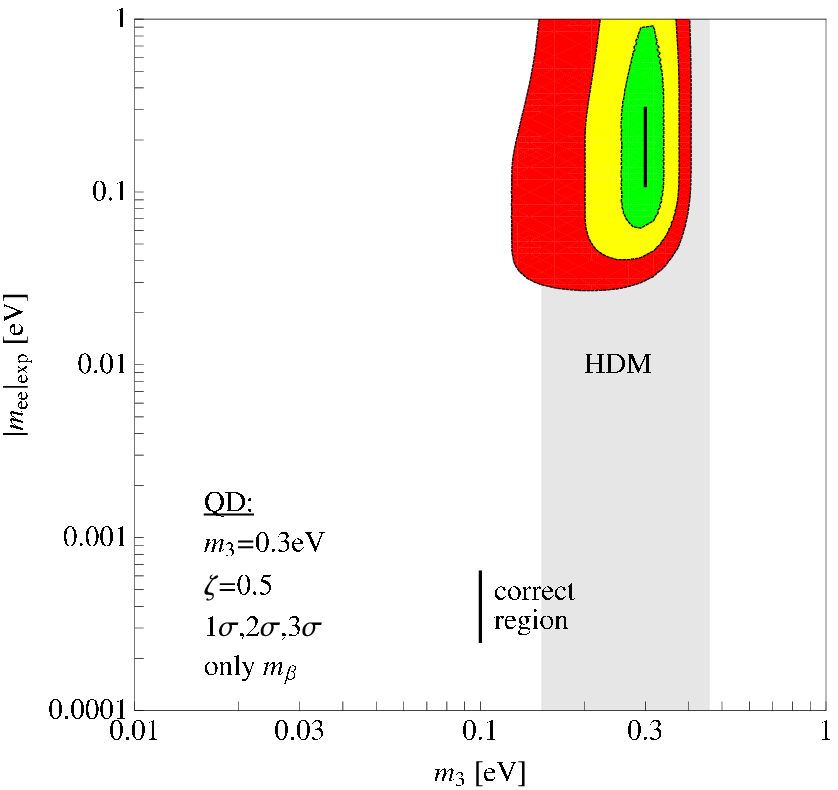,height=5.35cm,width=5.2cm}
\caption{\label{fig:chi2QD} 1$\sigma$, 2$\sigma$ and 
3$\sigma$ regions in the $m_3$-$\meff_{\rm exp}$ plane for 
the ${\cal QD}$ scenario. The left column shows the correct 
(solid line) as well as two possible incorrect cosmological 
measurements (dashed lines). The less desirable case, 
namely only taking into account a KATRIN measurement, 
is shown in the plots on the right. The 
area denoted HDM is the range of \meff~from the claim of 
part of the Heidelberg-Moscow collaboration.}
\end{center}
\end{figure}
\begin{figure}[t]
\begin{center}
\epsfig{file=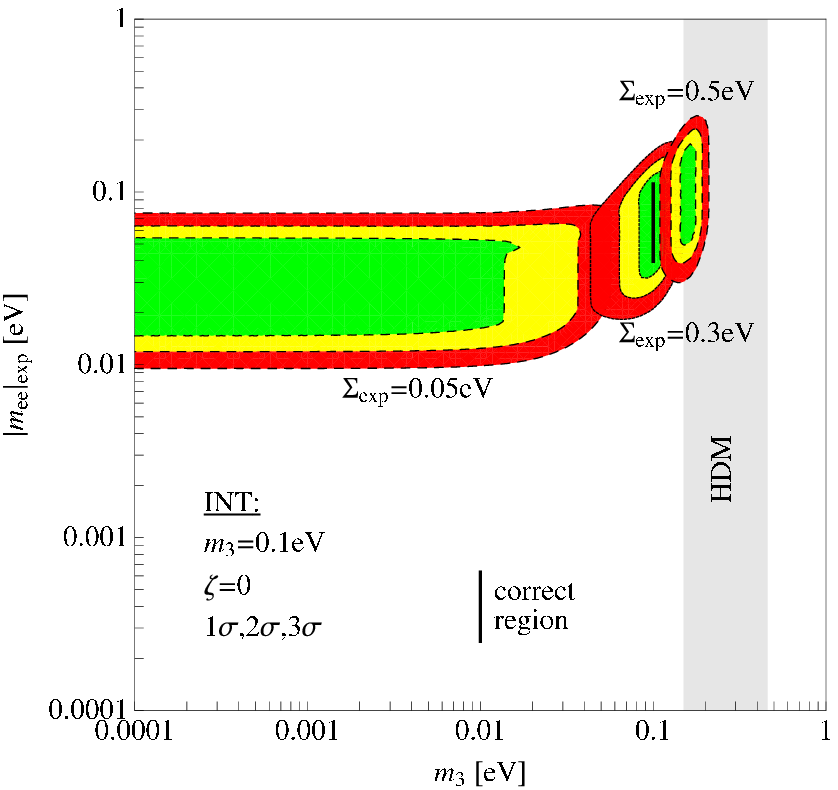,height=5.35cm,width=5.2cm}
\epsfig{file=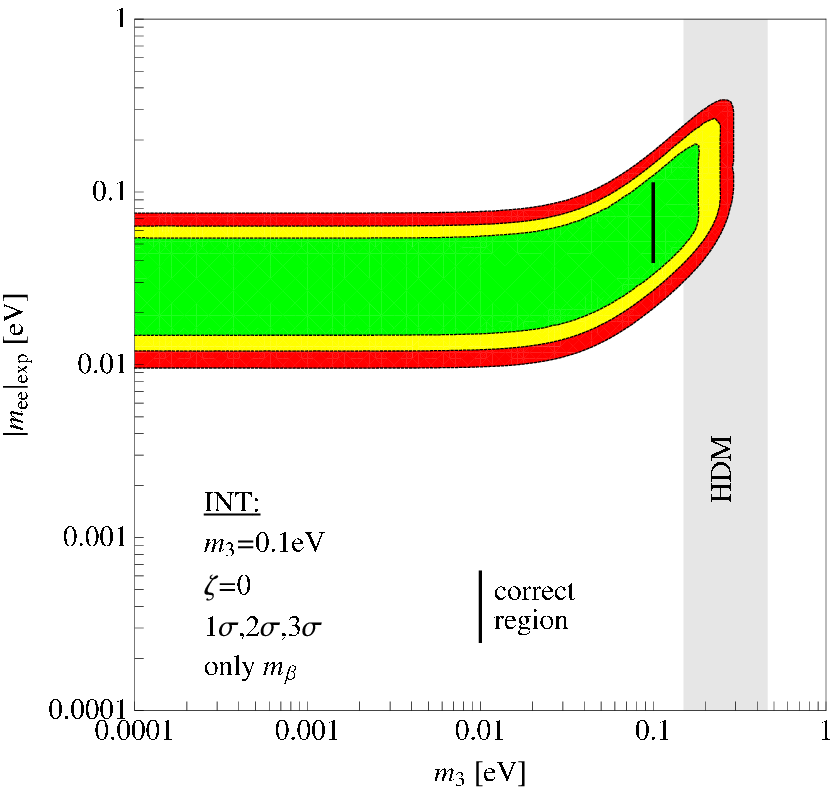,height=5.35cm,width=5.2cm}
\epsfig{file=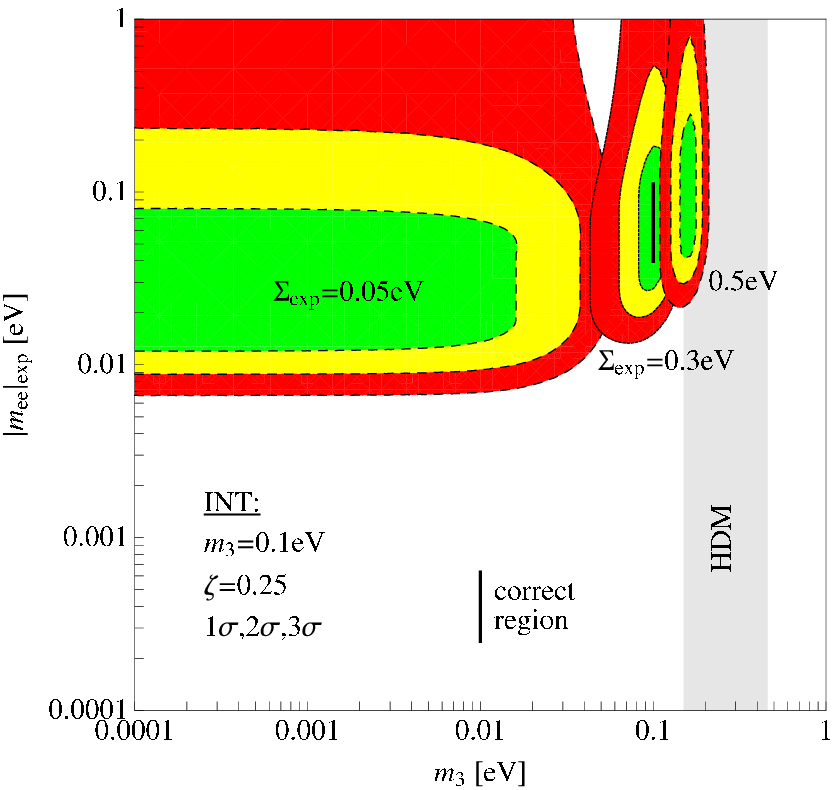,height=5.35cm,width=5.2cm}
\epsfig{file=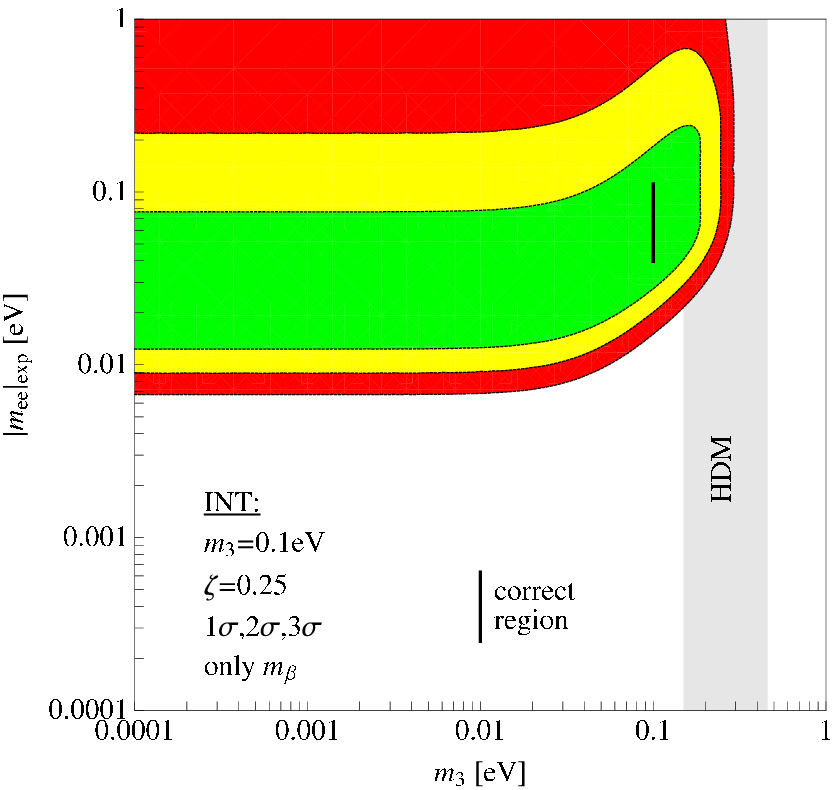,height=5.35cm,width=5.2cm}
\epsfig{file=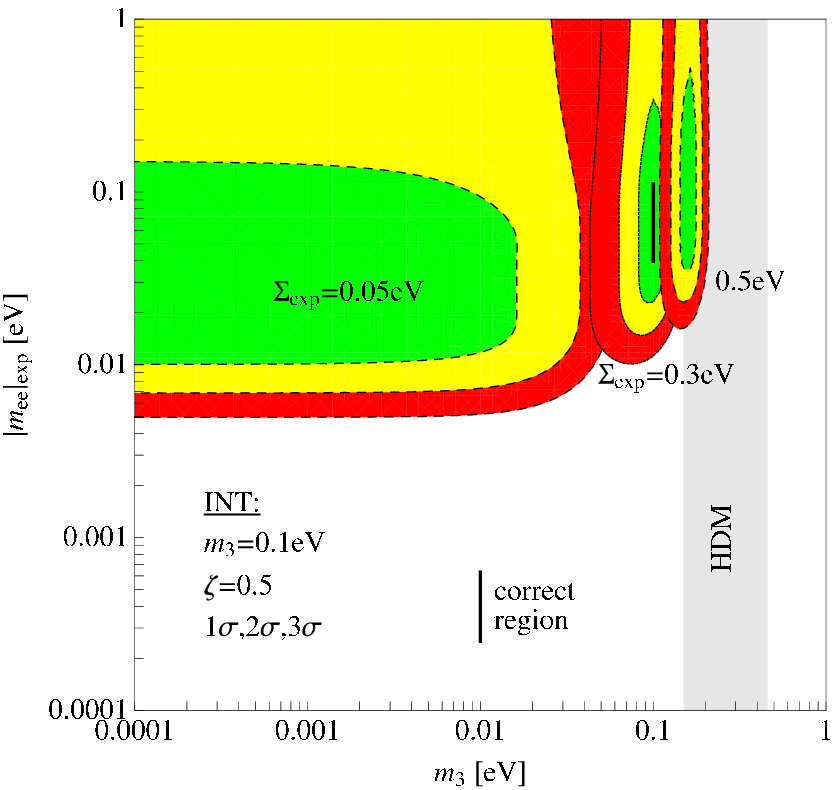,height=5.35cm,width=5.2cm}
\epsfig{file=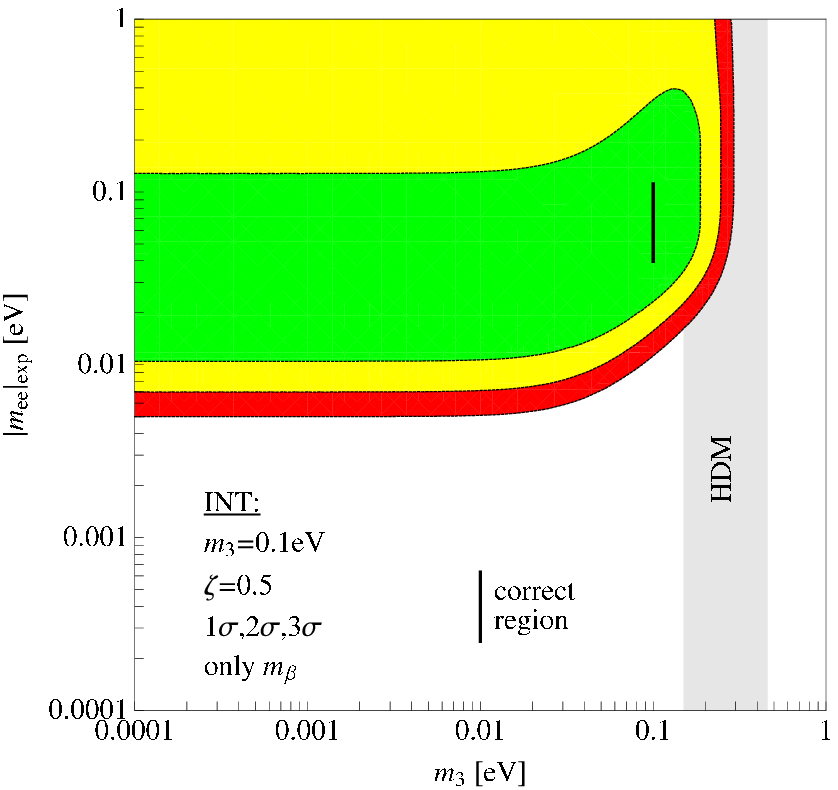,height=5.35cm,width=5.2cm}
\caption{\label{fig:chi2INT}Same as fig.~\ref{fig:chi2QD} for the ${\cal INT}$ scenario.}
\end{center}
\end{figure}
\begin{figure}[t]
\begin{center}
\epsfig{file=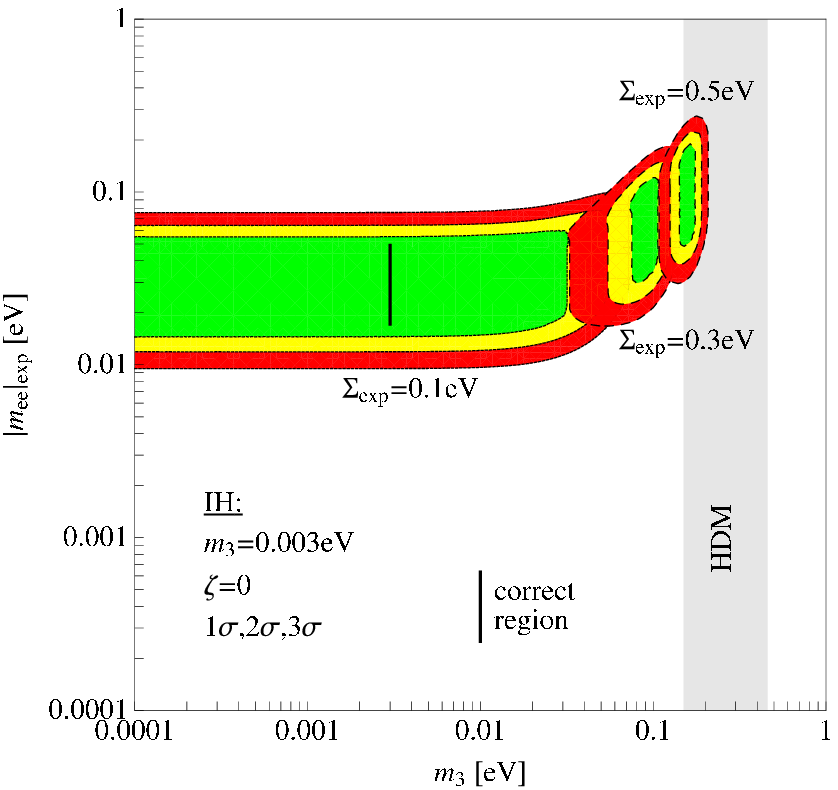,height=5.35cm,width=5.2cm}
\epsfig{file=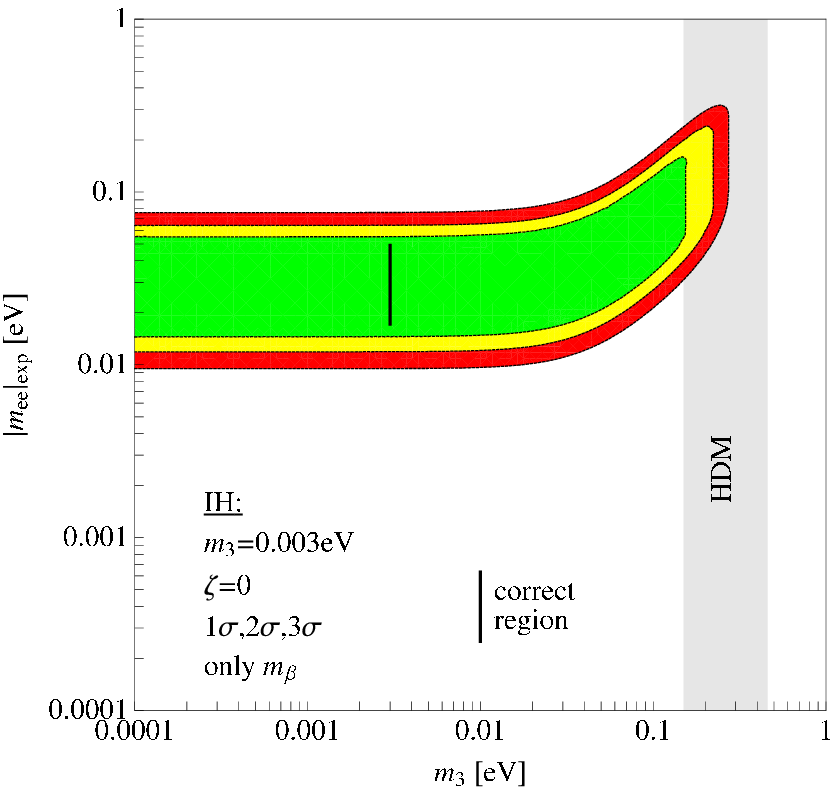,height=5.35cm,width=5.2cm}
\epsfig{file=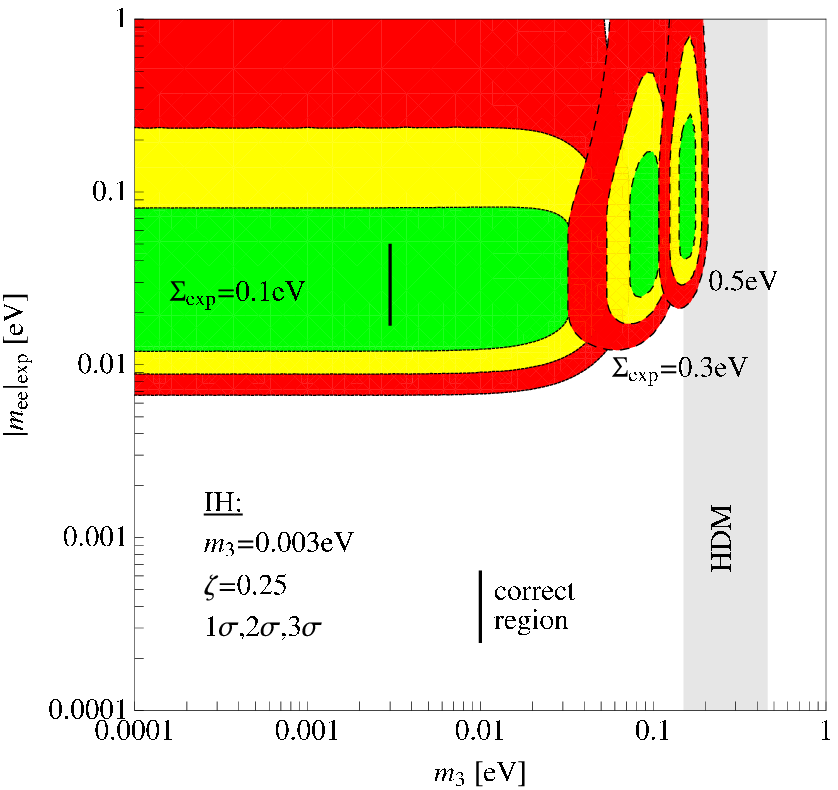,height=5.35cm,width=5.2cm}
\epsfig{file=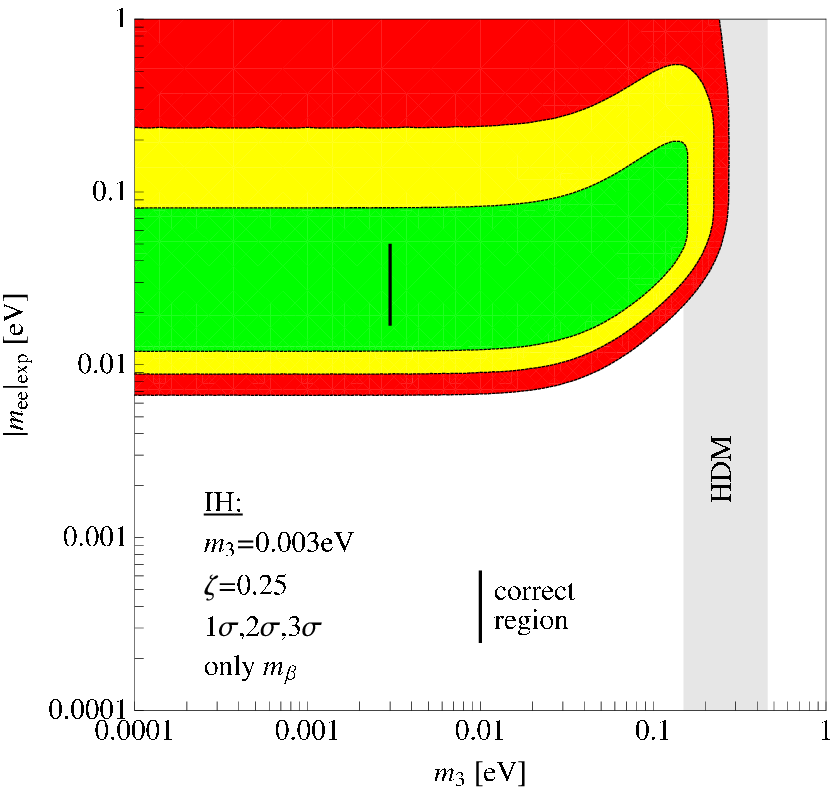,height=5.35cm,width=5.2cm}
\epsfig{file=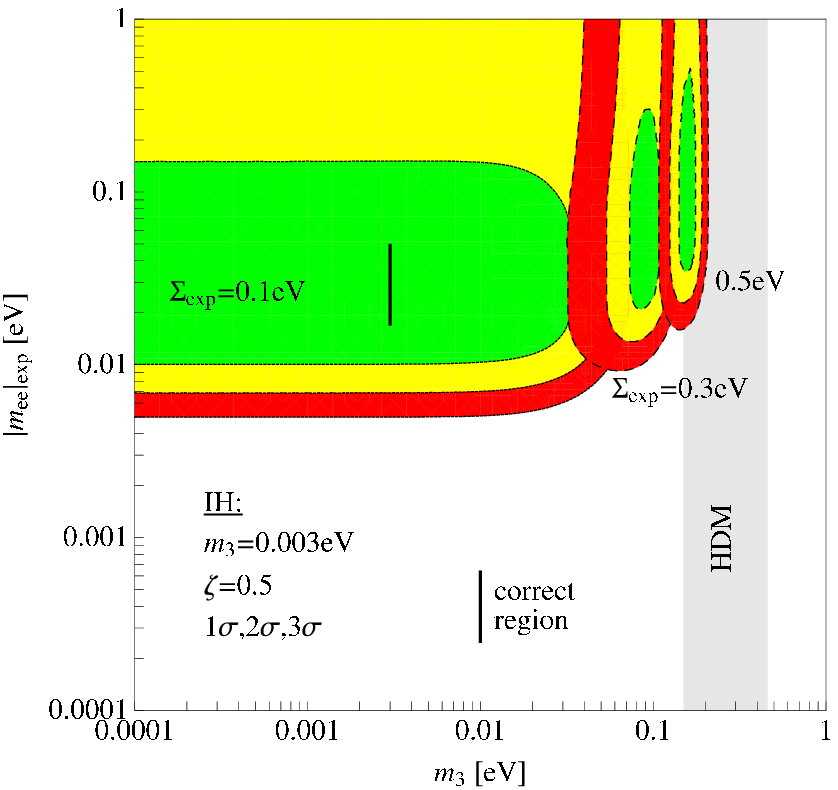,height=5.35cm,width=5.2cm}
\epsfig{file=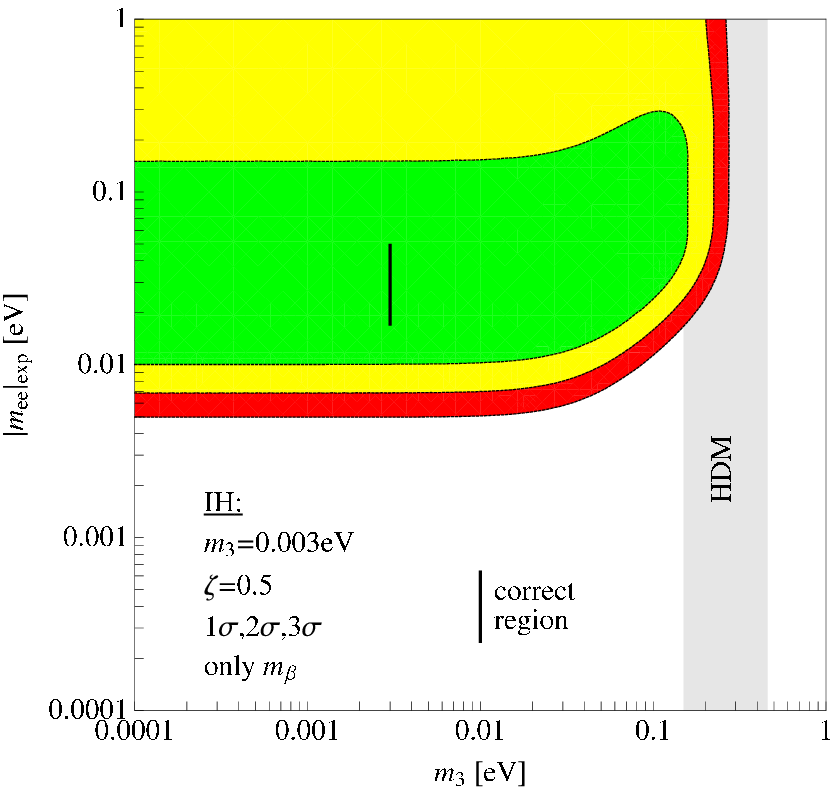,height=5.35cm,width=5.2cm}
\caption{\label{fig:chi2IH} 
Same as fig.~\ref{fig:chi2QD} for the ${\cal IH}$ scenario.}
\end{center}
\end{figure}

%{\bf 
The results of our analysis are shown as the 
solid lines in the left column of 
figs.~\ref{fig:chi2QD}, \ref{fig:chi2INT} and~\ref{fig:chi2IH}. 
In all cases, we have calculated the result for a 
consistent measurement (i.e., $m_\beta$ and $\Sigma$ are 
measured at their true values in the corresponding scenarios). 
The NME uncertainties we have chosen are $\zeta = 0$ 
(no uncertainty), 0.25 and 0.5. 
We have checked that values of $\zeta > 0.5$ 
will lead to results not too much different from the ones for 
$\zeta = 0.5$. 
The value $\zeta = 0.25$ is a quite 
typical one, cf.~refs.~\cite{nme,nme1}. 
This uncertainty arises from the highly non-trivial  
calculations of the nuclear part of the 
\onbb process. Different methods, and even different Ans\"atze 
within the same framework, differ in their result, and their spread 
is commonly taken into account as ``theoretical uncertainty''. 
Glancing at Fig.~5 in ref.~\cite{Simkovic:2009pp}, where the 
results of different methods of the NME calculation are compared 
for different nuclei including $^{76}$Ge, 
one can indeed see that the spread of the respective values 
around their mean value is about 0.2. 
We conclude that the values we use are realistic and typical.%} 

The true 
values of \meff~and $m_3$ are marked by the vertical black lines. 
The plots illustrate how well we can reconstruct the 
different scenarios for the various values of NME uncertainty. 
Having a look at fig.~\ref{fig:chi2QD}, we see that the 
${\cal QD}$ scenario can be 
reconstructed quite well, which is not surprising since in that case 
the KATRIN experiment as well as the cosmological measurement 
will provide a non-trivial signal. E.g., 
for $|m_{ee}|_{\rm exp}=0.20$~eV, the 1, 2 and 3$\sigma$ ranges for 
$m_3$ are $0.28-0.32$ eV, $0.27-0.33$ eV 
and $0.25-0.35$ eV, while the true value 
is 0.30~eV. Therefore, the reconstruction is quite accurate. This remains true also if the uncertainty in NME is non-zero because the plots are 
still narrow around the true value of $m_3$ (the numerical values 
suffer nearly no change) even though, with a larger 
NME uncertainty, also higher values of $|m_{ee}|_{\rm exp}$ 
are plausible. This is true for all three scenarios 
under consideration. 

Similar statements hold for the ${\cal INT}$ scenario shown in 
fig.~\ref{fig:chi2INT}, even though $m_\beta$ 
cannot be measured now. However, because there will still be 
a measurement of $\Sigma$, we have sufficient information 
on the neutrino mass.  In case the central measured value is 
$\meff_{\rm exp}=0.08$~eV and $\zeta=0$ the ranges are 
$0.08-0.12$~eV at 1$\sigma$ and $0.05-0.15$~eV at 3$\sigma$. 
In case of $\zeta=0.5$ we find $0.08-0.12$~eV at 1$\sigma$ and 
$0.04-0.15$ eV at 3$\sigma$. The mass scale has now a $3\sigma$ 
uncertainty of 50 \%, to be compared with roughly 15 \% in the 
${\cal QD}$ scenario. 

For ${\cal IH}$, in turn, there is no measurement 
that gives information $m_3$. Hence, it is 
only possible to give an upper limit on the smallest neutrino mass, 
as illustrated by the long horizontal band in the left 
column of fig.~\ref{fig:chi2IH}. Note that this band corresponds 
to the yellow band marking the inverted mass 
ordering in the upper plot of fig.~\ref{fig:1}. 
This upper limit is almost 
trivial, i.e., it corresponds to the neutrino mass limit 
obtainable from \obb~alone. To give some numerical values, for $\meff_{\rm exp}=0.04$~eV one would have the 
1~(3)$\sigma$ ranges $m_3 < 0.03~(0.07)$~eV 
for $\zeta=0$ and for $\zeta=0.5$. Due to the bound on 
$\Sigma$, there is very little dependence on $\zeta$.\\

Up to now, the discussion has focused on the case in which all 
measurements are compatible. As an example for inconsistency 
we discuss here a possible clash between results from 
KATRIN and from cosmology. 
To this end we leave $(m_\beta)_{\rm exp}$ equal to the true value 
of the corresponding scenario 
(new physics is not expected to influence $m_\beta$ \cite{NPmb}) 
and take values of $\Sigma_{\rm exp}$ which are smaller or 
larger than the true value. There are many scenarios or models 
in the literature which can lead to wrong values of $\Sigma$, 
see, e.g., refs.~\cite{CosmoKramer}. 
The result is shown by the areas within the 
dashed lines in the left columns of 
figs.~\ref{fig:chi2QD}-\ref{fig:chi2IH}. 
Having a look at ${\cal QD}$ first, we realize immediately that 
the physical range is reconstructed incorrectly. 
Hence, if there are systematic errors in the 
cosmological measurement, or unknown features in 
cosmology which we are not aware of, a wrong neutrino mass 
is reconstructed. In the ${\cal QD}$ case there is
still information from KATRIN, which leads to a reconstructed 
neutrino mass at most one order away from the true value, even if 
the wrong $\Sigma$ is taken into account. 
For the ${\cal INT}$ scenario, however, there is no information 
from KATRIN. Consequently, it might be that a {\it wrong} upper 
limit on $m_3$ is concluded, as illustrated by 
the long band for $\Sigma_{\rm exp}=0.05$~eV in the upper 
left plot of fig.~\ref{fig:chi2INT}. This is an example 
wherein one could draw a wrong conclusion by taking the 
cosmological measurement at face value. As expected, even worse 
cases may exist for the ${\cal IH}$ scenario. E.g., in 
the upper left plot 
of fig.~\ref{fig:chi2IH} one would, for 
$\Sigma_{\rm exp}=0.3$~eV, reconstruct a smallest neutrino 
mass of roughly 0.1~eV, to be compared with the true value of 
$m_3=0.003$~eV. 
For the ${\cal IH}$ scenario, one might not even realize that 
there is an inconsistency, since in that case, the 
KATRIN experiment can only provide an upper limit which 
is too far away from the true value of $m_3$.

One possible cross-check (or the possible consequence if one 
indeed finds that the results from KATRIN and from cosmology 
do not fit together) would be to dismiss the cosmological 
data altogether. We have also analyzed this case. 
Here, $S_{ab}$ from eq.~(\ref{eq:cov-matrix}) as well as 
$v_a$ would change from 3-dimensional to 2-dimensional objects while 
the rest of the procedure remains the same. The results 
for this analysis are plotted in the right columns of 
figs.~\ref{fig:chi2QD}-\ref{fig:chi2IH}, 
again for different values of the NME uncertainty. 
For ${\cal QD}$, the most optimal scenario, neglecting 
cosmology, would simply increase the errors in the 
determination of $m_3$: e.g., for $\meff_{\rm exp}=0.20$~eV and 
$\zeta=0$ the ranges are $0.26-0.34$~eV at 1$\sigma$ and 
$0.16-0.41$~eV at 3$\sigma$, while for $\zeta=0.5$ 
we find $0.26-0.34$~eV at 1$\sigma$ and $0.13-0.41$~eV at 3$\sigma$. 
The NME uncertainty has now a slightly bigger impact, and 
the error on $m_3$ increases by a factor of three, since now it 
is about 50 \% while it was roughly 15 \% when $\Sigma$ has been included 
in the analysis. 
For the ${\cal INT}$ scenario, 
however, there is a major difference to the former case: 
since now there is no other measurement besides $|m_{ee}|_{\rm exp}$ providing 
information on $m_3$, we can only derive an 
upper limit instead of determining a certain range for $m_3$. 
This is indicated by the band in the upper right plot of 
fig.~\ref{fig:chi2INT}. Finally, for ${\cal IH}$, the limit on 
$m_3$ gets only slightly worse compared to the case of a $\Sigma$, 
which is too small to be measured. In this case 
there would not even be a real drawback in taking into 
account the KATRIN result only. It remains to be said that 
in all cases a higher uncertainty for the NME does not significantly 
modify the conclusions in what concerns the value of $m_3$. 
%{\bf 
Finally, it is worth mentioning that if in ${\cal QD}$ 
scenarios the error on $\Sigma$ is decreased (increased), 
the obtained error on the neutrino mass is decreased (increased)   
by approximately the same factor.%}

With our analysis we can also compare the compatibility 
of our three benchmark scenarios with the range 
for $m_3$ of $0.15-0.46$ eV, calculated as the (global fit) 
$2\sigma$ range in ref.~\cite{Fogli:2006yq} from the claim in 
ref.~\cite{ev}. 
We give the implied range for $m_3$ 
as the gray band in figs.~\ref{fig:chi2QD}, \ref{fig:chi2INT} 
and~\ref{fig:chi2IH}. We see that scenario ${\cal QD}$ is 
consistent with the claim, even for a measurement of $\Sigma = 0.6$ 
eV, to be compared with the true value $\Sigma = 0.9$ eV. The 
${\cal INT}$ scenario (${\cal IH}$ scenario) 
is barely (very) incompatible for measured ``true'' 
values, but a too high value of $\Sigma_{\rm exp}$ can lead again 
to compatibility. We see that testing the claim and comparing 
it with cosmology is a non-trivial task (see also \cite{nme1}).

\section{\label{sec:3}Conclusions}
In this work we have investigated possible constraints on the 
neutrino mass in future experiments. We assumed realistic errors 
on the observables, in particular for neutrino-less double 
beta decay. Then, we have checked how certain realistic 
benchmark scenarios, which correspond to different regimes for the 
smallest neutrino mass, can be reconstructed from future 
measurements. Furthermore, we have 
pointed out how wrong conclusions could be drawn from inconsistent 
results, i.e., if cosmology provides a wrong value for the sum of 
neutrino masses.  
In case of consistent measurements we may summarize as follows: 
typical $3\sigma$ errors for quasi-degenerate 
neutrino masses range from roughly 15 \% (including $\Sigma$) 
to 50 \% (excluding $\Sigma$), where NME uncertainties 
play a larger role in the latter case. Intermediate scale masses 
can also be determined with 50 \% uncertainty. In case of an inverted hierarchy, the effective mass is constant for a large range of the smallest mass, which allows only to derive upper limits on it.

%%%%%%%%%%%%%%%%%%%%%%%%%%%%%%%%%%%%%%%%%%%%%%%%%%%%%%%%%%%%%%%%%%%%%%%
\section*{Acknowledgments}
%%%%%%%%%%%%%%%%%%%%%%%%%%%%%%%%%%%%%%%%%%%%%%%%%%%%%%%%%%%%%%%%%%%%%%%
We are grateful to T.~Schwetz for valuable discussions. 
This work was supported by the ERC under the Starting Grant 
MANITOP (W.R.) and by the Deutsche Forschungsgemeinschaft 
in the Transregio 27, as well as 
by the EU program ILIAS N6 ENTApP WP1.

\end{document}